\documentclass[preprint,12pt,sort&compress]{elsarticle}
\usepackage{amsmath,amssymb}

\usepackage{graphicx}
\usepackage{epstopdf}
\usepackage{color}
\usepackage{siunitx}
\newcommand{\beq}{\begin{equation}}
\newcommand{\eeq}{\end{equation}}

\journal{arXiv}

\begin{document}
\begin{frontmatter}

\title{On mechanisms of electromechanophysiological interactions between the components of nerve signals in axons}
\author{J\"uri Engelbrecht, Kert Tamm, Tanel Peets}
\address{Laboratory of Solid Mechanics, Department of Cybernetics, School of Science,\\ Tallinn University of Technology, Akadeemia tee 21, Tallinn 12618, Estonia, \\E-mails: je@ioc.ee, kert@ioc.ee, tanelp@ioc.ee}

\begin{abstract}
Recent studies have revealed the complex structure of nerve signals in axons. There is experimental evidence that the propagation of an electrical signal (action potential) is accompanied by mechanical and thermal effects. In this paper, first an overview is presented on experimental results and possible mechanisms of electromechanophysiological couplings which govern the signal formation in axons. This forms a basis for building up a mathematical model describing an ensemble of waves. Three physical mechanisms responsible for coupling are (i) electric-lipid bi-layer interaction resulting in the mechanical wave in biomembrane; (ii) electric-fluid interaction resulting in the mechanical wave in the axoplasm; (iii) electric-fluid interaction resulting in the temperature change in axoplasm. The influence of possible changes in variables which could have a role for interactions are analysed and the concept of internal variables introduced for describing the endothermic processes. The previously proposed mathematical model is modified reflecting the possible physical explanation of these interactions. 
\end{abstract}

\end{frontmatter}

\section{Introduction}

Although the studies into the propagation of signals in nerves have a long history \citep{Bishop1956,Nelson2004}, the research is going on. It is accepted that beside the electrical signal (action potential), the accompanying non-electrical manifestations like mechanical effects together with local temperature changes should also be taken into account in order to get a full picture on this fascinating phenomenon. There is strong experimental evidence \cite{Iwasa1980,Tasaki1988,Tasaki1989,Yang2018,Terakawa1985,Ritchie1985,Howarth1968} on accompanying effects and there are several mathematical models \cite{Engelbrecht2018d,Tamm2019,Heimburg2005,Chen2019,Hady2014} proposed for describing a signal with such effects. However, there is no general consensus about the possible physical mechanisms governing the processes of propagation and coupling. A need for such a consensus is stressed by many researchers, from Hodgkin \citep{Hodgkin1964a} to recent studies \citep[etc.]{Andersen2009,Drukarch2018}. In this paper, a brief overview on experimental results and proposed physical mechanisms of coupling is given and analysed within the framework of a robust mathematical model \citep{EngelbrechtMEDHYP,Engelbrecht2018e}. As a result, on the basis of known experimental studies, the coupling forces are specified in more detail in order to reflect better the physics of the process. It is concluded that there are several mechanisms responsible for emerging an ensemble of waves in nerve fibres. Based on experimental results, these mechanisms are cast into the mathematical descriptions.

In Section~2 a brief overview of experimental studies is given. The possible physical mechanisms are analysed in Section~3. On the basis of this analysis, possible mathematical descriptions and coupling forces are proposed in Section~4. The mathematical model with coupling forces related to physical effects is described in Section~5 which enlarges our previous proposals \citep{EngelbrechtMEDHYP,Engelbrecht2018e,Tamm2019}. Finally, the conclusions are presented in Section~6.

\section{Brief overview of experimental studies}

The contemporary understanding of processes in nerve fibres is based on experimental studies in axons. In simplified physical terms, axons are cylindrical tubes embedded into an extracellular fluid. The wall of the tube has a bilayered lipid structure called biomembrane formed by phospholipids (a head group and a fatty acid tail) which altogether is about 3-4~nm thick. Inside the tube is axoplasm (intracellular fluid) with cytoskeletal elements where the action potential (AP) propagates. The squid giant axon which is the classical experimental object has the diameter up to 1~mm but the usual mammalian axons have the diameter around 20~$\si{\um}$. A more detailed description of axon morphology is given, for example, by Debanne et al.~\citep{Debanne2011} and of biomembranes -- by Mueller and Tyler \citep{Mueller2014}.

The way to understanding the propagation of APs is paved by experiments of Hodgkin and Huxley \citep{Hodgkin1952}. They have measured the AP in the non-myelinated squid axon and explained the role of ionic (Na$^+$ and K$^+$) currents. The absolute amplitude of the AP was about 100~mV, duration (without the overshoot) about 1~ms and the velocity 18.9~ms$^{-1}$. The contemporary measurements have demonstrated that the velocities of electric signals in nerves vary in a large interval (from ca 1~ms$^{-1}$ to ca 100~ms$^{-1}$). The measurements of the AP together with the quantitative measurements of ionic currents \citep{Hodgkin1952} form the basis for the corresponding Hodgkin-Huxley (HH) model \citep{Hodgkin1964a} which is nowadays called also the Hodgkin-Huxley paradigm.

Many experiments have shown that the propagation of an AP in a nerve fibre is accompanied by transverse displacements of the biomembrane which mean changes in the axon diameter \citep{Iwasa1980,Tasaki1988,Tasaki1989}. These local changes called also swelling are small being in the range of 1-2~nm and compared with the diameter of fibres are of several orders smaller. This is confirmed by recent experiments by Yang et al.~\citep{Yang2018}. In addition to the deformation of the biomembrane, the pressure wave in axoplasm has been measured by Terakawa \citep{Terakawa1985}. In a squid axon the amplitude of a pressure wave, measured simultaneously with the AP, was about 1 to 10~mPa. The temperature change during the passage of an AP is also measured, it is of the range about 20-30~$\si{\micro}$K for garfish \citep{Tasaki1988} and much less for bullfrog \citep{Tasaki1992}. The earlier findings on mechanical and thermal effects are summarised by Watanabe \citep{Watanabe1986}, and more recently by Andersen et al.~\citep{Andersen2009}.

The transverse displacement $W$ of the cylindrical biomembrane is associated to the longitudinal displacement $U$ of the biomembrane. This effect -- $W$ is proportional to the gradient of $U$ -- is well understood in mechanics \citep{Porubov2003} for the theory of rods. It means that the bipolar $W$ measured by Tasaki \citep{Tasaki1988} corresponds to the unipolar $U$ (and vice versa). The possible deformation of a biomembrane under loading is studied by measuring the transverse displacement \citep{Gonzalez-Perez2016,Perez-Camacho2017} and interpreted then as an accompanying mechanical wave along the biomembrane \citep{Heimburg2005,Gonzalez-Perez2016}. This longitudinal wave may have a soliton-type shape \citep{Heimburg2005}, i.e., is unipolar. It must be noted that the excitable plant cells (\emph{Chara braunii}) behave in a similar way: the electrical signal is coupled with a mechanical effect \citep{Fillafer2018}.

To sum up, there is strong experimental evidence about several effects accompanying the propagation of APs. We shall analyse next the proposed mechanisms of coupling.  

\section{Possible physical mechanisms of coupling}

The whole ensemble of waves and effects is the following: (i) action potential AP and the corresponding voltage $Z$ associated with the ion current(s), here denoted by $J$; (ii) longitudinal wave LW in the biomembrane with the amplitude $U$; (iii) pressure wave PW in the axoplasm with the amplitude $P$; (iv) transverse displacement TW with the amplitude $W$; (v) temperature change $\Theta$. Note that above just one ion current is listed as a variable. The crucial question is: what are the physical mechanisms which link these signal components into a whole.

The starting point of the analysis is related to causality. The classical understanding in axon physiology is the HH paradigm: the whole process is electricity-centred and starts with generating the AP. However, according to the original HH model, the accompanying effects (see above) are not considered.  Recently much attention is also paid to another paradigm: the whole process starts with generating the mechanical disturbance (LW) as proposed by Heimburg and Jackson \citep{Heimburg2005}. The governing equation of the LW has a soliton-type solution describing the region of higher density in the biomembrane and that is why this approach is called soliton theory. The pros and contras of both approaches are analysed by Appali et al.~\citep{Appali2010} and Meissner \citep{Meissner2018a}.
 
Later in Section 4 we follow the HH paradigm as the fundamental approach in contemporary axon physiology \citep{Clay2005,Debanne2011} but try to link it to the accompanying effects. This means that one should pay attention to electrical-to-mechanical and mechanical-to-electrical couplings and to possible heat generation.


\subsection*{Qualitative observations from experiments}
The following observations can be noted from the published experiments:
-The measured TW is bi-polar for the squid giant axon and its peak coincides with the peak of the AP \citep{Tasaki1988}, both have approximately the same duration.\\
-The measured TW is close to uni-polar for the garfish olfactory nerve and its peak coincides with the peak of the AP \citep{Tasaki1989}.\\
-The peak of the force developed at the axon surface coincides fairly accurately with the peak of the AP \citep{Tasaki1988}.\\
-The peak of the pressure PW lags behind the peak of the AP for the squid giant axon \citep{Terakawa1985}.\\
  - Mechanical and thermal signals are in phase with the voltage changes \citep{Gonzalez-Perez2016}.\\
   - The shape and the width of a TW are similar to those of the measured AP (without the overshoot) for the rat neuron \citep{Yang2018}.\\
    - The experiments with \emph{Chara braunii} cells have demonstrated that the mechanical pulse (out-of-plane displacement of the cell surface) propagates with the same velocity as the electrical pulse and is (in most cases) of bi-polar nature \citep{Fillafer2018}.\\
    - The AP and temperature $\Theta$ for the garfish olfactory nerve are almost in phase and the duration of the positive phase of heat production is very close to the duration of the depolarizing phase of the AP \citep{Tasaki1988}.\\
     - The AP is narrower than the temperature change and the thermal response cannot be directly proportional to the change of voltage \citep{Ritchie1985}.\\
      - Good correlation exists between the initial positive heat and the potassium (K$^+$) leakage \citep{Howarth1968}.
      
\subsection*{Possible mechanisms}
The well-studied HH model describes the AP dependence on ion currents \citep{Hodgkin1964a}. Based on experiments with the squid giant axons, the role of opening and closing of ion channels has been demonstrated resulting in the asymmetric shape of the AP with an overshoot, the existence of a threshold for generating the pulse and the refractory period. The HH model is analysed in many details as the fundamental model of axon physiology \citep{Debanne2011}. It has been shown that in addition to Na$^+$ and K$^+$ ion currents there may be many more ion currents regulating the AP properties, especially for human atrial APs \citep{Courtemanche1998,Bean2007}. However, the HH model does not describe accompanying effects, although Hodgkin said \citep{Hodgkin1964a}: ``In thinking about the physical basis of the action potential perhaps the most important thing to do at the present moment is to consider whether there are any unexplained observations which have been neglected in an attempt to make the experiments fit into a tidy pattern''. 

The physical mechanism governing the HH model is based on the flow of ions through the biomembrane upon the change of voltage in axoplasm. Contemporary understanding is that the ion channels may be voltage-gated like in the HH model but also mechanically-sensitive \citep{Mueller2014,Ranade2015}. This must be taken into account in building up a fully coupled model.

Gross et al.~\citep{Gross1983} have analysed electromechanical transductions in nerves. For electrical-to-mechanical transductions the mechanisms of electrostriction and piezoelectricity are analysed and argued that both mechanisms could predict the swelling effects. For mechanical-to-electrical transductions it is proposed that the stress-induced changes due to surface charges influence the intracellular electric field.

A promising mechanism for coupling the electrical and mechanical signals is the flexoelectric effect which is manifested in the deformation of the biomembrane curvature under an imposed electric field \citep{Petrov2006}. The flexoelectric effect is used by Chen et al.~\citep{Chen2019} for modelling the coupling of the AP and mechanical wave (TW). The classical HH model combined with cable theory includes density change in the biomembrane induced by the flexoelectricity. The biomembrane is modelled as an elastic (or viscoelastic) tube where the flexoelectric force is included into the governing equation. This force depends on the local change in the membrane potential. The changes in the axon diameter are taken into account and the system is reciprocal -- it can be triggered either by an electrical pulse resulting in an AP or by a mechanical stimulus.

A coupled model based on the primary AP which generates all other effects is proposed by El Hady and Machta \citep{Hady2014}. A Gaussian profile (a pulse) for an AP is taken as a basic without modelling. The assumption is made that in the fibre the membrane has potential energy, and axoplasmic fluid has kinetic energy.  The idea is that a surface wave (meaning the surface of the fibre) is generated in the membrane and in the bulk field within the axon the linearised Navier-Stokes equations are used for calculating the pressure.  The ensemble includes voltage (pulse, not a typical AP), radial membrane displacement TW and the lateral displacement inside the axon (i.e., the PW). The heat is assumed to be produced as additional release of mechanical energy, summing transverse changes in the diameter and lateral stretch. So the sequence: AP -- mechanical waves -- heat (temperature) is followed. 

Rvachev \citep{Rvachev2010} has proposed that the axoplasmic pressure pulse (PW) triggers all the process. The PW triggers the Na$^+$ channels and the local HH voltage spike develops which in its turn opens the Ca$^{2+}$ channels. Free intracellular Ca$^{2+}$ activates then the contraction of filaments in the axoplasm which gives rise to the radial contraction in the lipid bilayer. The similar idea that a PW could cause the excitement is proposed by Barz et al.~\citep{Barz2013}.

Based on experimental results, Terakawa \citep{Terakawa1985} has suggested that the pressure PW arises either from a change in electrostriction across the axoplasm or from a change in charge-dependent tension along the axoplasm. He states that pressure response is correlated with membrane potential and not with the membrane current. A slight influence of electro-osmotic water flow to pressure response is detected.

Abbott et al.~\citep{Abbott1958} discuss the heat generation in Maia nerves. They give three possible reasons for heat production: (i) the positive heat is derived from the energy released during the rising phase of the AP and the negative heat due to the absorption of energy during the falling phase of the AP; (ii) the positive heat is due to the interchange of Na$^+$ and K$^+$ ions and the negative heat represents the partial reversal of this interchange; (iii) the heat production is related to exothermic and endothermic chemical reactions. Richie and Keynes \citep{Ritchie1985} have supported similar experimental data like Abbott et al.~\citep{Abbott1958}. They also stated that energy of the membrane capacitor is proportional to the voltage square and that the thermal response cannot be directly proportional to the change of voltage.

Tasaki and Byrne \citep{Tasaki1992} have analysed the heat production in bullfrog myelinated nerve fibres. They estimated theoretically the relation of voltage $V$ to temperature $\Theta$ in dependence of time constant $RC$ (capacity x resistance in operational amplifier). For $RC$ longer, they used $V$ related to $\Theta$ and for $RC$ shorter $V$ related to $d\Theta/dt$. 

To sum up, there is no consensus about the coupling and transduction of energy. Some studies concentrate upon the electromechanical transductions, some – upon the coupling of an electrical signal and temperature. This way or another, the qualitative experimental observations serve as guides in modelling. According to the HH paradigm, the process is triggered by an electrical stimulus which generates the AP but it is also proposed that the stimulus could be of a mechanical character. This could be the LW \citep{Heimburg2005} or the PW \citep{Rvachev2010}. The model by Chen et al.~\citep{Chen2019} based on using the flexoelectric effect involves reciprocity of electrical and mechanical components of the process. The generation of the temperature during the process is associated either with the electrical signal \citep{Abbott1958,Tasaki1992} or to mechanical effects \citep{Hady2014}. However, the role of chemical reactions in producing temperature changes is also under discussion \citep{Abbott1958}.

This way or another, in most of studies (both experimental and theoretical), the coupling is associated to local changes of fields which cause changes in the whole system. 

\section{Modelling of coupled signals and coupling forces}

In order to overcome (or to unite) the possible differences in proposing the various mechanisms of interactions described in Section~3, the best way is to return to basics. This means starting from physical considerations. Indeed, the signal propagation in nerves is a dynamical process which in terms of continuum theories is far from the molecular range (although some molecular mechanisms may be of influence) and the amplitudes of the components (AP, PW, TW, $\Theta$) are in micro or meso-scale. The question is then what is the role of wave equations or diffusion equations which usually govern the physical dynamical processes, in nerve pulse propagation. The celebrated HH model is based on the telegraph equation where the inductivity is neglected but ion currents added resulting in a reaction-diffusion equation which predicts a finite velocity of the AP. Leaving aside the physiological details, the longitudinal wave in the biomembrane and the pressure wave in axoplasm are waves and the role of wave equations describing the process is obvious. And in physical systems where temperature effects are measured, the Fourier law and the diffusion equation are the cornerstones of the possible description. So the conservation laws, well-known in continuum theories could be used. In general formulation they all include the possible forces which correspond to given physical situations (see, for example, \citep{Eringen1962}) and may also include more variables for which some thermodynamical constraints must be satisfied \citep{Engelbrecht2015b}. The next step is to describe the forces in correspondence with governing mechanisms. Although for the modelling of signals in nerves, several mechanisms (see Section~3) have been proposed, it is a challenge to find the proper physical description of possible forcing. Certain flexibility is desirable because it has been mentioned in several studies (for example,  in \citep{Abbott1958}) that there might be several mechanisms involved simultaneously in the signal propagation.

Consequently, the crucial issue is how to construct the forces and how to model the coupling. It is proposed that the main hypothesis for constructing these forces could be \citep{EngelbrechtMEDHYP}: all mechanical waves in axoplasm and surrounding biomembrane together with the heat production are generated due to changes in electrical signals (AP or ion currents) that dictate the functional shape of coupling forces. The seconding hypotheses are: the changes in the pressure wave may also influence the waves in biomembrane and mechanical waves may influence the AP and ion currents. This means reciprocity between the signal components. In many studies referred to in Section~3, the changes are mentioned \citep{Gross1983,Chen2019,Terakawa1985,Tasaki1992} as reasons for interactions. And back to the history: the German physiologist Emil Du Bois-Reymond has noticed in the 19th century that ``the variation of current density, and not the absolute value of the current density at any given time, acts as a stimulus to a muscle or motor nerve'' \citep{Hall1999}. This statement is called the Du Bois-Reymond law. 

What is change? In mathematical terms changes mean either space ($X$) or time ($T$) derivatives of variables. This gives the glue for proposing the functional shapes of forces which at the first approximation could be described in the form of first-order polynomials of gradients or time derivatives of variables ($Z_X$, $J_X$, $U_X$, $P_X$, and $Z_T$, $J_T$, $U_T$, $P_T$). Here and further subscripts $X$  and $T$ denote partial derivatives with respect to space and time, respectively. Such an approach involves also certain flexibility in choosing the model. 

Further on, a possible approach in modelling is envisaged based on ideas described above \citep{EngelbrechtMEDHYP,Engelbrecht2018e,Tamm2019}. The processes which compose the leading effects in signal propagation can each be described by single model equations. In the coupled model these single equations are united into a system by coupling forces. For a proof of concept, the coupling could be modelled by a simpler approach: from the AP and ion currents to all the other effects. It has been proposed that the process could be divided into primary and secondary components \citep{Engelbrecht2018arXiv}. The primary components are characterised by corresponding velocities and their mathematical models are derived from wave equation(s). These components are the AP, LW and PW. The secondary components are either derived from the primary components like TW or their models are derived from basic equations which do not possess velocities like temperature $\Theta$. In the latter case the diffusion-type equation could serve as a basic mathematical model.

On the basis of experimental studies (Section~3) it seems plausible that there are several physical mechanisms of coupling. This concerns electrical-mechanical (AP to PW and AP to LW) and electrical-thermal (AP to $\Theta$) transduction. In essence, the mechanisms should also include feedback and coupling between all the components of the ensemble. However, these effects could be of more importance in pathological situations (axon dysfunction) or for detailed understanding of neural communication and neural activity in general.

As stated above, the coupling forces could be described by changes in variables. Note that gradients (space derivatives) act along the axon and time derivatives across the membrane. Based on thermodynamics (see Richie and Keynes \citep{Ritchie1985}, Tamm et al, \citep{Tamm2019}), for temperature changes one should consider also the possible effects of $Z$ or $Z^2$  (alternatively $J$ or $J^2$).
In general terms the basic mechanisms might be the following:\\
(i) electric-biomembrane interaction resulting in a mechanical response (LW, variable $U$);\\
(ii) electric-fluid (axoplasm) interaction resulting in a mechanical response (PW, variable $P$);\\
(iii) electric-fluid (axoplasm) interaction resulting in a thermal response ($\Theta$).\\

However, as said above, the feedback from mechanical waves to other components of the whole ensemble might also influence the process. In more detail these interactions can be characterised:\\
(a) influence from the AP: 
\begin{enumerate}
	\item 	Pressure change in axoplasm -- $Z_X$, i.e., AP gradient along the axon axis can influence the pressure as a result of  charged particles present in axoplasm reacting to the potential gradient along the axon;
	\item Pressure change in axoplasm -- $Z_T$, i.e., potential changes across the biomembrane can lead to a pressure change proportional to the potential change through electrically motivated membrane tension changes (also taken into account for LW, and this is accounting for the influence from the mechanical displacement which is taken proportional to $Z_T$);
	\item 	Pressure change in axoplasm -- $J_T$, i.e., the axoplasm volume change from ion currents in and out of axon through the biomembrane plus the effect of possible osmosis which is assumed to be proportional to the ionic flows;
	\item 	Density change in biomembrane -- $Z_T$, i.e., electrically induced membrane tension change or flexoelectricity; 
	\item Density change in biomembrane -- $J_T$, i.e., membrane deformation as a result of an ionic flow through the membrane (ion channels deforming surrounding biomembrane when opening/closing); note that ion channels are not modelled here in the FHN model explicitly; 
	\item 	Temperature change --  $Z^2$ or $J^2$, i.e.,  temperature increase from current flowing through the environment (power); this effect is related to Joule heating;
	\item  Temperature change -- endothermic term is dependent on the integral of $J$    which is taken to be proportional to concentration of reactants which decays
    exponentially in time after the signal passage; needs a new kinetic equation to 
    be added. 
\end{enumerate}
(b) influence from the pressure P:
\begin{enumerate}
	\item Density change in biomembrane -- $P_T$, i.e., membrane deformation (displacement) from the local pressure changes inside the axon; 
	\item Temperature change -- $P_T$, i.e., reversible local temperature change: when pressure increases then temperature increases proportionally and when pressure decreases then temperature decreases proportionally, this happens at the same timescale as pressure changes;
	\item 	Temperature -- the irreversible local temperature increase from energy consumed by viscosity (friction) -- this is actually time integral of $P_T$.
\end{enumerate}
(c) influence from the mechanical wave U in biomembrane:
\begin{enumerate}
	\item Action potential change -- $U_T$, i.e., ion currents are suppressed when membrane density is increased and amplified when membrane density is decreased;
	\item Temperature change -- $U_T$, i.e., the reversible local temperature change, when density increases the local temperature is increased proportionally and when density decreases the local temperature decreases proportionally, this happens at the same timescale as the density changes in biomembrane;
	\item Temperature change --  the irreversible local temperature increase from energy consumed by the added friction/viscosity term in longitudinal density change model -- this is actually time integral of $U_T$ which is proportional to $U$.
\end{enumerate}

These are the plausible processes which can influence the dynamics of nerve pulse propagation. However, not all of the listed effects might be energetically at the same scale as far as dynamics of the nerve pulses are concerned. At this stage, the possible influence of temperature changes might have on properties of fibres are disregarded as experimental observations are showing temperature changes low enough (up to about 50 $\mu$K) to be probably negligible for these processes \cite{Ritchie1985,Tasaki1992}. In addition, the possible changes in the fibre diameter which could change the physical properties are also disregarded being of many scales lower compared with leading effects. In general, the significance of the listed effects must be determined by experimental and theoretical studies in the future.


\section{Mathematical model with specified coupling forces}
We have proposed a mathematical model of nerve pulse propagation including the accompanying effects \citep{EngelbrechtMEDHYP,Engelbrecht2018c}. This model served as a proof of concept and was based on using the basic equations of mathematical physics which were modified in order to reflect the physiological effects. The following models have been used: the reaction-diffusion model for describing the AP (the FHN model) and accompanying ion current; the modified wave equation for describing the LW in the biomembrane (the iHJ model); the wave equation with dissipation for the PW; the diffusion equation for the $\Theta$. The coupling forces \citep{EngelbrechtMEDHYP} have been proposed to model the interaction between the components of the whole ensemble. The criticism on this model \cite{Holland2019} is based on the seeming inconsistency of the coupling of the AP and LW described by indicated models. The main point of the criticism is that in the HH model the capacitance of the neural membrane is taken constant  \cite{Holland2019}. However, it must be noted that in the coupled model one has used the modelling of an AP only for obtaining a correct (measured by numerous experiments) asymmetric shape of an AP with an overshoot and a corresponding ion current. Note that the overshoot has a definite role in signal propagation \citep{Andersen2009,Debanne2011}. Certainly, other models could also be used for this purpose. For example, El Hady and Machta have used just a symmetrical Gaussian pulse without an overshoot for the AP \citep{Hady2014}. The present understanding of electrophysiology give strong evidence of the importance of electrical signals \citep{Clay2005,Debanne2011} in axons with an asymmetric shape. 

In mathematical terms, such ``simple conceptual models can be used to gain insight, develop intuition, and understand ``how something works'''' \citep{Wooley2005}. In physical terms, much must be understood before one could realise a complete unifying model. Following the HH paradigm \citep{Noble2010}, the causality relies on the propagation of an electrical signal in the axon. 
It must be noted that, alternatively, following Heimburg and Jackson \citep{HJ2007} or Mussel and Schneider \citep{Schneider2018} the mechanical wave in a biomembrane can generate voltage changes. 
The manifestations of physical processes described briefly in Section~3 are suggesting the causality related to the AP but indicate the need to modify  both the model equations as well as coupling forces for describing the accompanying effects. The main reason for the modification is the possible plurality of dynamical processes and coupling effects. Changes compared with the previous model \citep{EngelbrechtMEDHYP,Engelbrecht2018e,Tamm2019} are following: dissipation is added to the iHJ equation; the coupling force for the temperature equation is specified accounting for multiple mechanisms which include also endothermic processes. The latter means proposing an additional kinetic equation involving the ion current. It means that the whole system involves the dissipative effects and the endothermic processes involve internal variables. Internal variables are not observable and characterise the internal structure or processes \citep{Maugin1994a}. It has been used for describing some microscopic effect leading to local structural rearrangements \citep{Maugin1994a} which have no inertial influence. Such a concept is widely used in contemporary continuum mechanics \citep{Berezovski2017} and relies on thermodynamical description of dissipative effects. The first attempt to use internal variables in nerve pulse dynamics is described by Maugin and Engelbrecht \citep{Maugin1994} for describing the role of ion currents. Here such a formalism \citep{Van2008a} is used for describing the possible chemical processes \citep{Abbott1958} which influence the temperature changes. The mechanicsm of such changes is proposed to be similar to the behaviour of phenomenological variables in the HH model -- the change from one level to another.

For sake of completeness, the modified model is presented here. The FHN equations describing the AP are 
\beq\label{fhneq}
\begin{split}
Z_{T} &=  D Z_{XX} - J + Z \left( Z -  C_1  - Z^2 +  C_1  Z \right),\\
J_{T} &=  \epsilon_1 \left( C_2  Z - J \right),
\end{split}
\eeq
where $C_i = a_i + b_i$ and $b_i = -\beta_i U$. Here
$Z$ is action potential, $J$ is ion current, $\epsilon_1$ is the time-scales difference parameter, $a_i$ is the ``electrical'' activation coefficient, $b_i$ is the ``mechanical'' activation coefficient and $U$ is a longitudinal density change from lipid bi-layer density model. 

For the PW a modified wave equation with is used
\beq \label{peq}
P_{TT} = c_{2}^{2} P_{XX}  - \mu_2 P_T + F_2(Z,J),
\eeq
where $P$ is pressure, $\mu_2$ is a viscosity coefficient, $F_2$ is the coupling term accounting for the possible influence from the AP and TW.

Longitudinal wave in lipid bilayer is modelled by the improved HJ model
\beq \label{ihjeq}
\begin{split}
U_{TT} &=  c_{3}^{2} U_{XX} + N U U_{XX} + M U^2 U_{XX} +
   N U_{X}^{2} + 2 M U U_{X}^{2}\\ &-
   H_1 U_{XXXX} + H_2 U_{XXTT} - \mu_3 U_T + F_3(Z,J,P),
\end{split}
\eeq
where $U = \Delta \rho$ is the longitudinal density change, $c_{3}$ is the sound velocity in the unperturbed state, $N$,$M$ are nonlinear coefficients, $H_i$ are dispersion coefficients and $\mu_3$ is the dissipation coefficient. Note that $H_1$ accounts for the elastic properties of the bi-layer and $H_2$ the inertial properties. The $F_3$ is the coupling term accounting for the possible influence from the AP and PW. The transverse displacement (TW) is $W \propto U_X$. 

The local temperature change $\Theta$  is governed by
\beq \label{tempeq}
\Theta_{T} = \alpha \Theta_{XX} + F_4(Z,J,U),
\eeq
where $\Theta$ is the temperature change, $\alpha$ is a thermal conductivity coefficient and $F_4$ is the coupling term accounting for the possible influence from the AP and LW.

The coupling term $F_2$ is
\beq \label{f1eq}
F_2 =  \eta_1 Z_X + \eta_2 J_T + \eta_3 Z_T,
\eeq 
where $Z_X$ accounts for the presence of charged particles under the influence of the potential gradient (along the axon), 
$J_T$ accounts for the ionic flows into and out of an axon (across the membrane) and
$Z_T$ accounts for the possible pressure change as a result of membrane tension changes from the electrical field. 
Contact term $F_3$
\beq \label{f2eq}
F_3 =  \gamma_1 P_T + \gamma_2 J_T - \gamma_3 Z_T,
\eeq
where $P_T$ accounts for possible membrane deformation because of pressure changes (pressure to TW to LW), 
$J_T$ accounts for possible membrane deformation as a result of ionic flows through ion channels and
$Z_T$ accounts for possible electrically induced membrane tension change. Note the sign, assuming that density decreases with the increasing tension.

For $F_4$ the following values were used in \cite{Tamm2019}: $F_4 = \tau_1 Z$ or  $F_4= \tau_2 Z^{2}$ or $F_4 = \tau_3 J$ or  $F_4= \tau_4 J^{2}$  or $F_4 = \tau_5 U$ or  $F_4= \tau_6 U^{2}$ or $F_4 = \tau_7 Z_T + \tau_8 J_T$ or $F_4 = \tau_9 J_T + \tau_{10} U_X,$ etc. 
These terms were considered from the viewpoint of possible mathematical description. However, it is possible that Eq.~\eqref{tempeq} needs to be improved in order to match better experimental observations when the time-scale for the temperature changes is larger than the time-scale for AP changes. Note that Abbott et al.~\cite{Abbott1958} have argued about the possible combination of many reasons for heat production. Consequently, for modelling the temperature changes one could use
\beq \label{tempeq2}
\Theta_{T} = \alpha \Theta_{XX} + F_4(Z,J,P,U),
\eeq
where
\beq \label{f3eq}
F_4 = \tau_1 J^{2} + \tau_2 \left( P_T + \varphi_2(P) \right) + \tau_3 \left( U_T + \varphi_3(U) \right) - \tau_4 K,
\eeq
or
\beq \label{f3eqz}
F_4 = \tau_1 Z^{2} + \tau_2 \left( P_T + \varphi_2(P) \right) + \tau_3 \left( U_T + \varphi_3(U) \right) - \tau_4 K.
\eeq
Here $K$ is an internal variable, modelling all endothermic processes. In principle, it describes the change from one level to another like the phenomenological variables describing ion currents in the HH model \cite{Hodgkin1952}. The corresponding evolution equation for $K$ in Eqs~\eqref{f3eq} and \eqref{f3eqz} is \citep{Berezovski2017,Maugin1994}
\beq \label{keq}
K_T + \epsilon_4 K = \zeta J.
\eeq
An alternative way of representing the evolution of $K$ could be
\beq \label{keq2}
K_T = \varphi_4(J) - \frac{K-K_0}{\tau_K}, \quad \text{where} \quad K_0  = 0, \quad \tau_K = \frac{1}{\epsilon_4},
\eeq
where $K_0$ is equilibrium level (in our setting equilibrium has been taken as zero level) and $\tau_K$ is the relaxation time. Clearly, $K$ as an internal variable has a certain relaxation time before reaching its equilibrium value. Following the discussion of Abbott et al.~\citep{Abbott1958}, the thermal influence of chemical reactions might initially be endothermic and in the later stages of the recovery -- exothermic. 
In principle, this internal variable can describe both endo-- and  exothermic chemical influence, which means that the value of K has to change the sign (overshooting equilibrium value). However, this might need several ion currents to be taken into account.

In Eqs \eqref{f3eq} and \eqref{f3eqz}
\beq \label{f3PUint}
\varphi_2(P) = \lambda_2 \int P_T dT \quad \text{and} \quad \varphi_3(U) = \lambda_3 \int U_T dT, 
\eeq
where integrals in \eqref{f3PUint} characterise energy transfer from mechanical waves into the temperature increase as a result of friction--type terms in Eqs \eqref{peq} and \eqref{ihjeq} and $\lambda_i$ are coefficients. In dimensionless case $\lambda_i = \mu_i$. 

In Eqs~\eqref{f3eq} -- \eqref{keq} $\tau_i, \lambda_i, \zeta, \epsilon_i$ are coefficients. The term $J^2$ in Eq.~\eqref{f3eq} (or $Z^2$ in Eq.~\eqref{f3eqz}) is either related to $\mathcal{P} = J^2 r$ or $\mathcal{P} = Z^2/r$ where $r$ is resistance and $\mathcal{P}$ is power accounts for the Joule heating from the (electrical) current flowing through the environment. The term $P_T$ accounts for the local temperature increase if the pressure increases and decrease when pressure decreases (reversible), term $\int P_T dT$ accounts for the temperature increase from the energy lost to viscosity in Eq.~\eqref{peq}, term $U_T$ accounts for the local temperature increase when the local density is increased and decrease if the local density decreases (reversible), term $\int U_T dT$ accounts for the temperature increase from the energy lost to viscosity in Eq.~\eqref{ihjeq} and finally, as noted, the term $K$ takes into account the temperature decrease from endothermic processes. In Eq.~\eqref{keq}, term $J$ is used on assumption that endothermic process intensity is proportional to
\beq
\varphi_4(J) = \zeta \int J dT
\eeq
which is taken as the time integral of ion currents at the location. This concentration-like quantity is decaying exponentially in time which is accounted by the term $- \epsilon_4 K$ in Eq.~\eqref{keq}. In other words, we are considering some kind of endothermic chemical reaction. Some other processes are outlined in the literature \citep{Heimburg2008,Schneider2018} but not considered here, as possibly relevant, like, the phase change of the lipid bi-layer. In the present framework the phase change, if relevant, could be accounted by the term $U_T$ when using an alternate physical interpretation where change in the membrane density is a result of the phase change. However, unless noted otherwise the interpretation used in the present framework for the $U$ is that this is the `normal' density change as a result of mechanical wave propagating in elastic environment. 

\begin{figure}
\centering
\includegraphics[width=0.75\textwidth]{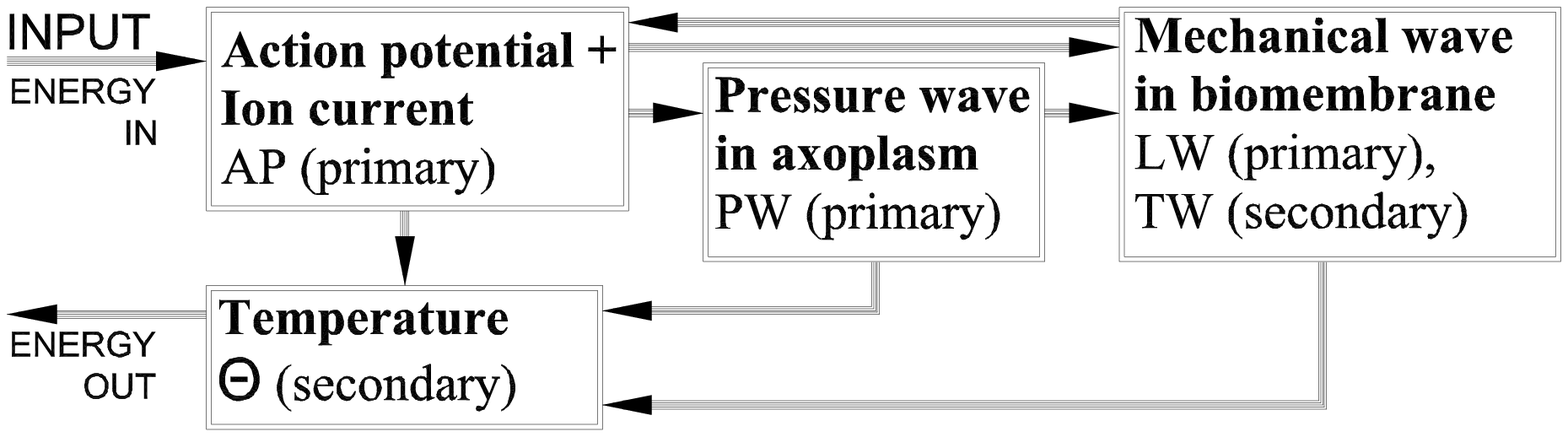}
\caption{Block diagram of the model system.}
\label{blokkskeem}
\end{figure}

A block diagram of the presented model is shown in Fig.~\ref{blokkskeem}, where the main components of the model framework  and interactions between previously established individual component models are highlighted. The proposed framework is highly flexible and any of the components can be replaced with an improved model with the desired descriptive detail, like, for example, one could use an HH model instead of the FHN model without having to change the framework significantly.

\begin{figure}
\includegraphics[width=0.33\textwidth]{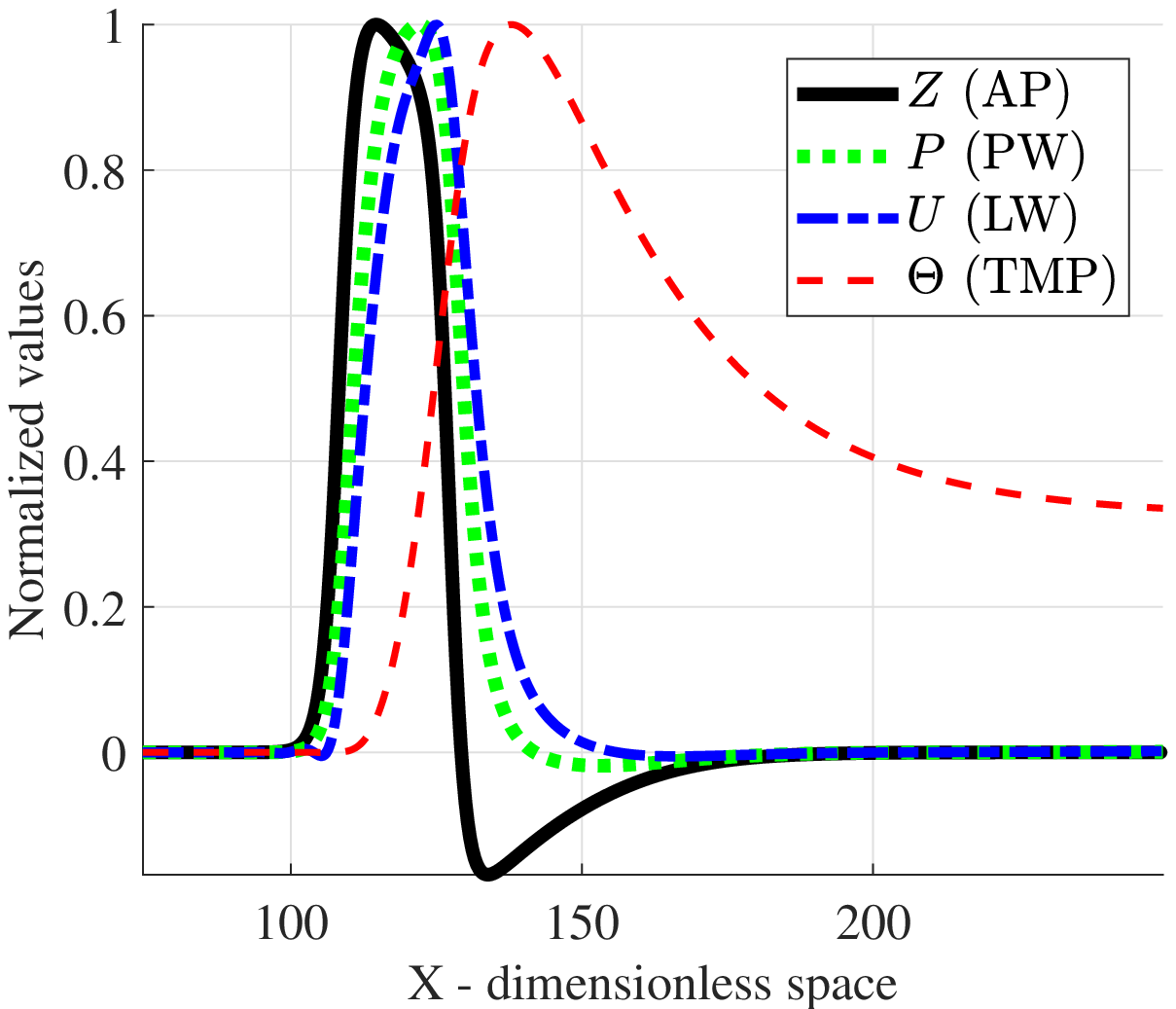}
\includegraphics[width=0.315\textwidth]{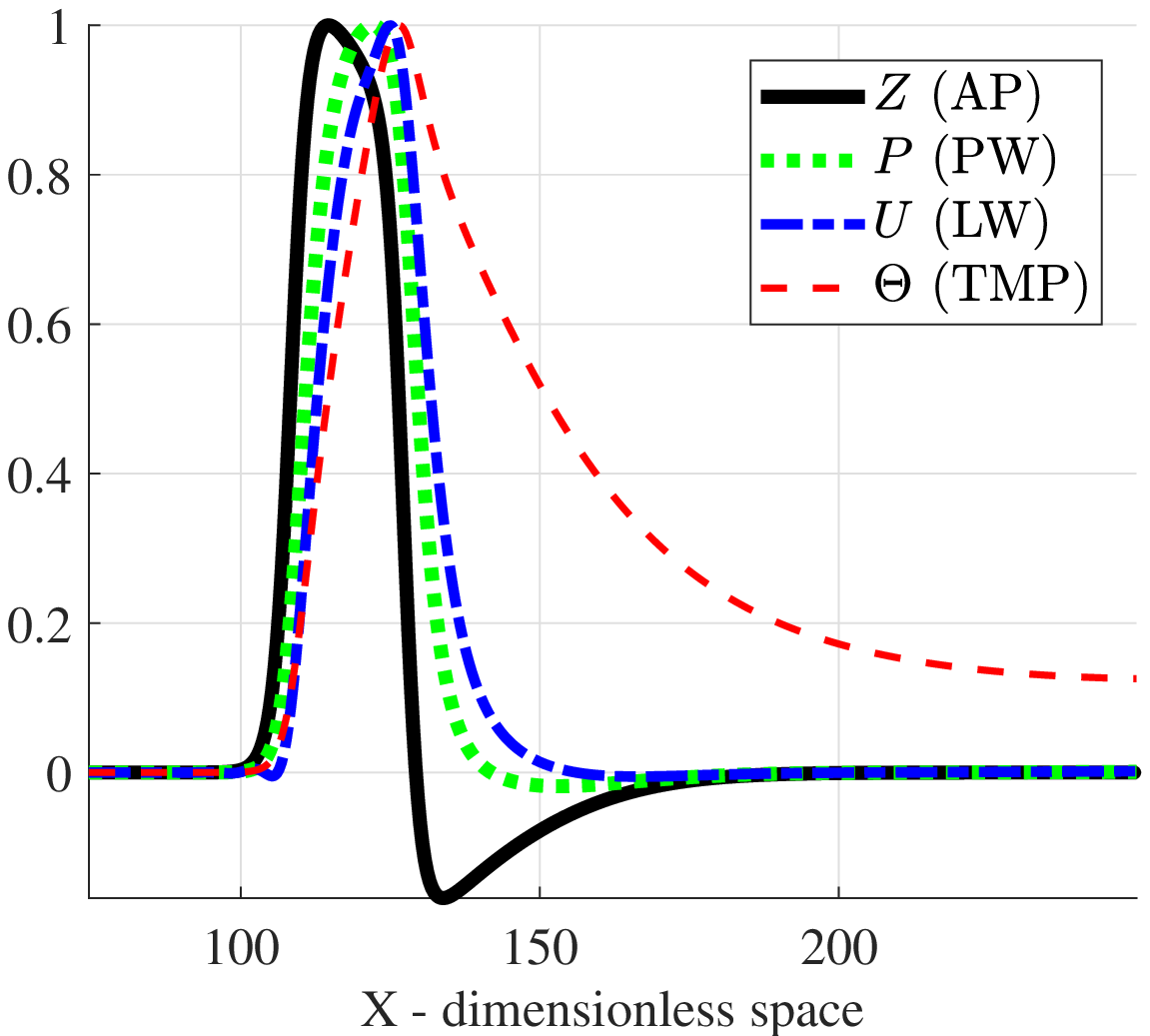}
\includegraphics[width=0.315\textwidth]{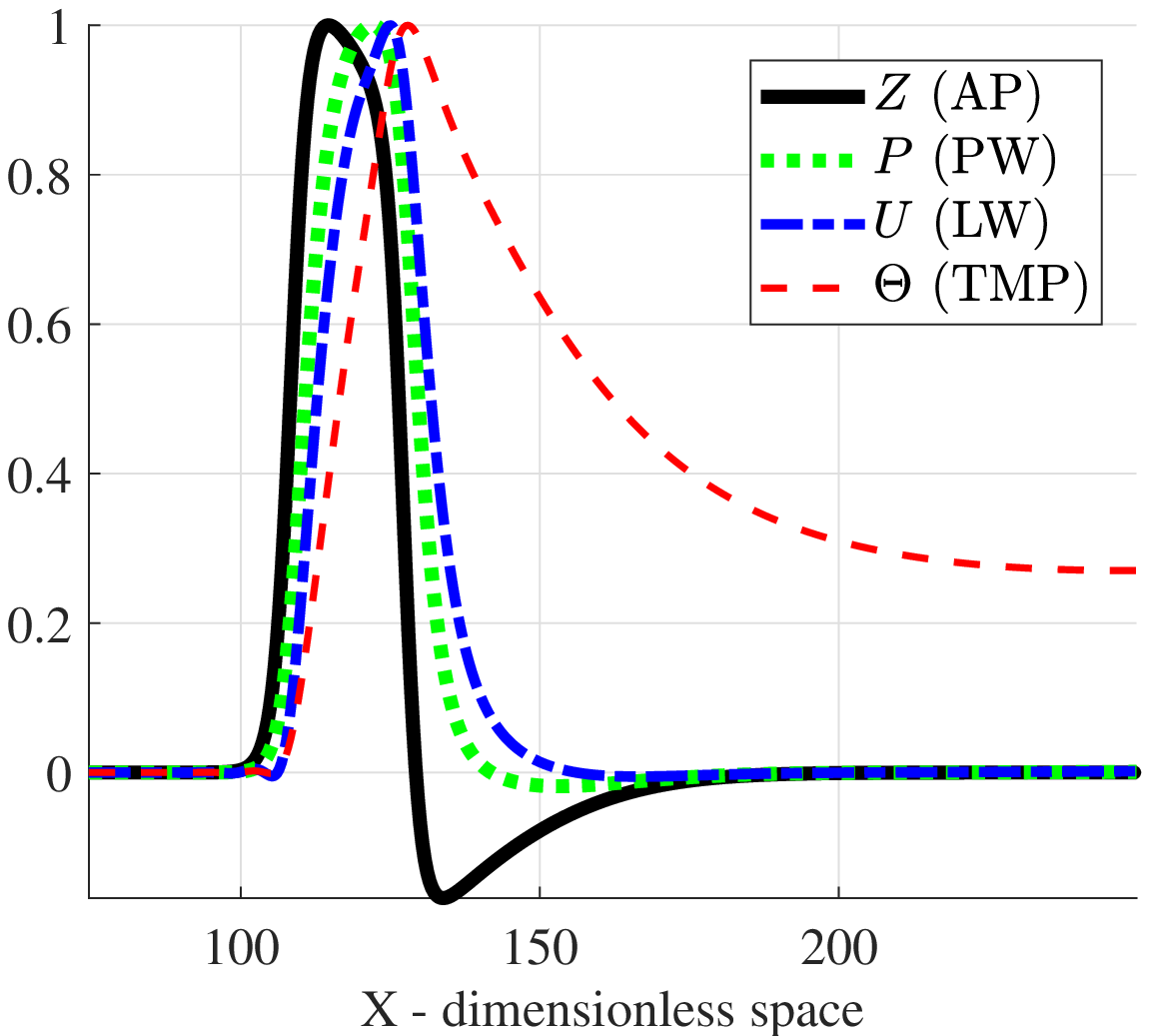}
\caption{Heat signals. Left panel (thermal source only $\propto J^2$) -- $\tau_1=0.05;\,\tau_2=0,\,\tau_3=0$, middle panel (thermal source only $\propto \varphi_2(P)$) -- $\tau_1=0;\,\tau_2=0.1,\,\tau_3=0$, right panel (thermal source only $\propto \varphi_3(U)$) -- $\tau_1=0;\,\tau_2=0,\,\tau_3=0.5$}\label{Fig1}
\end{figure}

Example results for the cases where  only a single source of thermal energy is present is shown in Fig.~\ref{Fig1}. The profiles for the $Z,P,U$ are the same but which is different is the thermal response $\Theta$ curves. Qualitatively what is relevant here is the location of the thermal response peak compared to the wave ensemble and the shape of the initial thermal response curve. 

\begin{figure}
\centering
\includegraphics[width=0.33\textwidth]{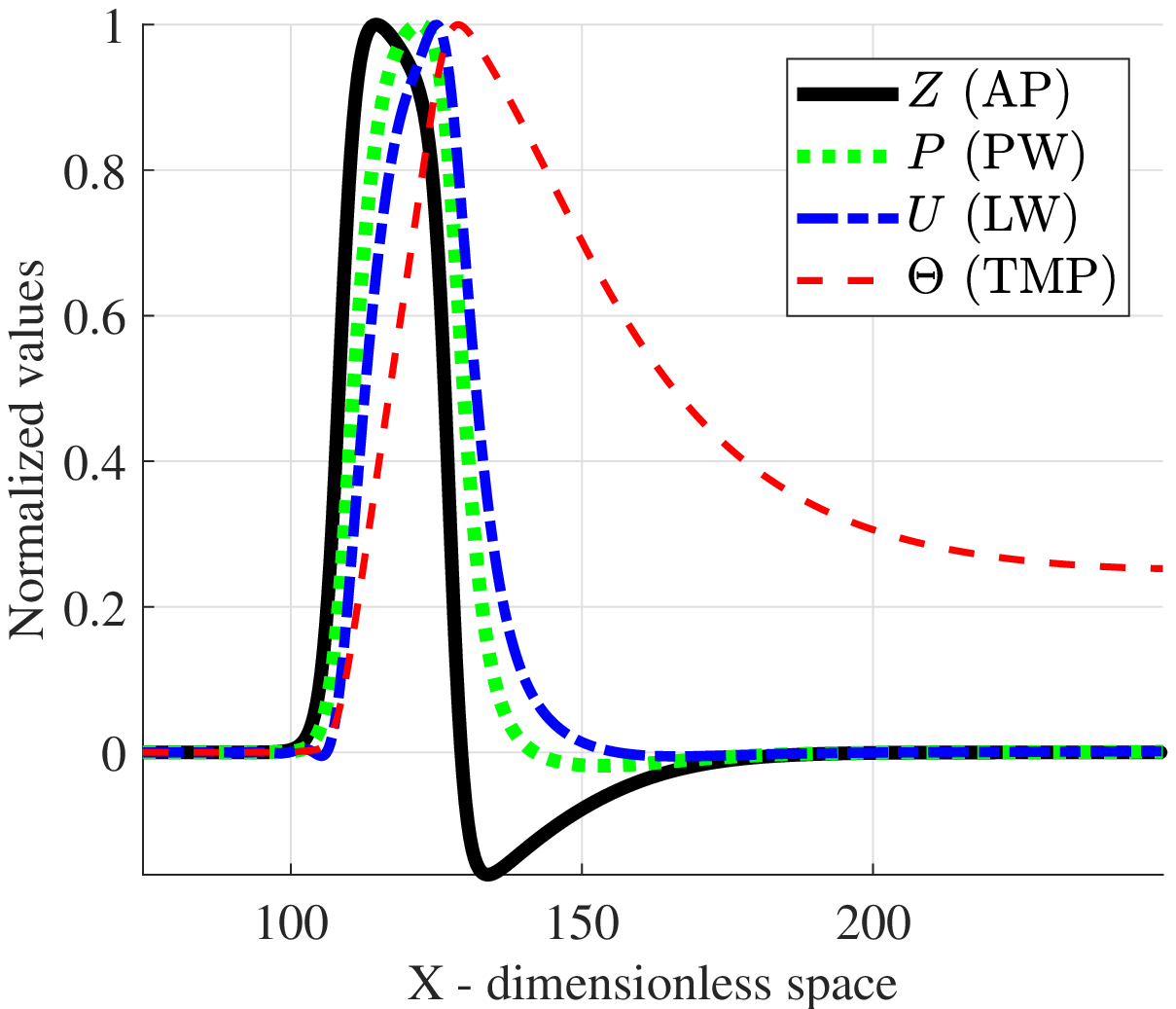}
\includegraphics[width=0.33\textwidth]{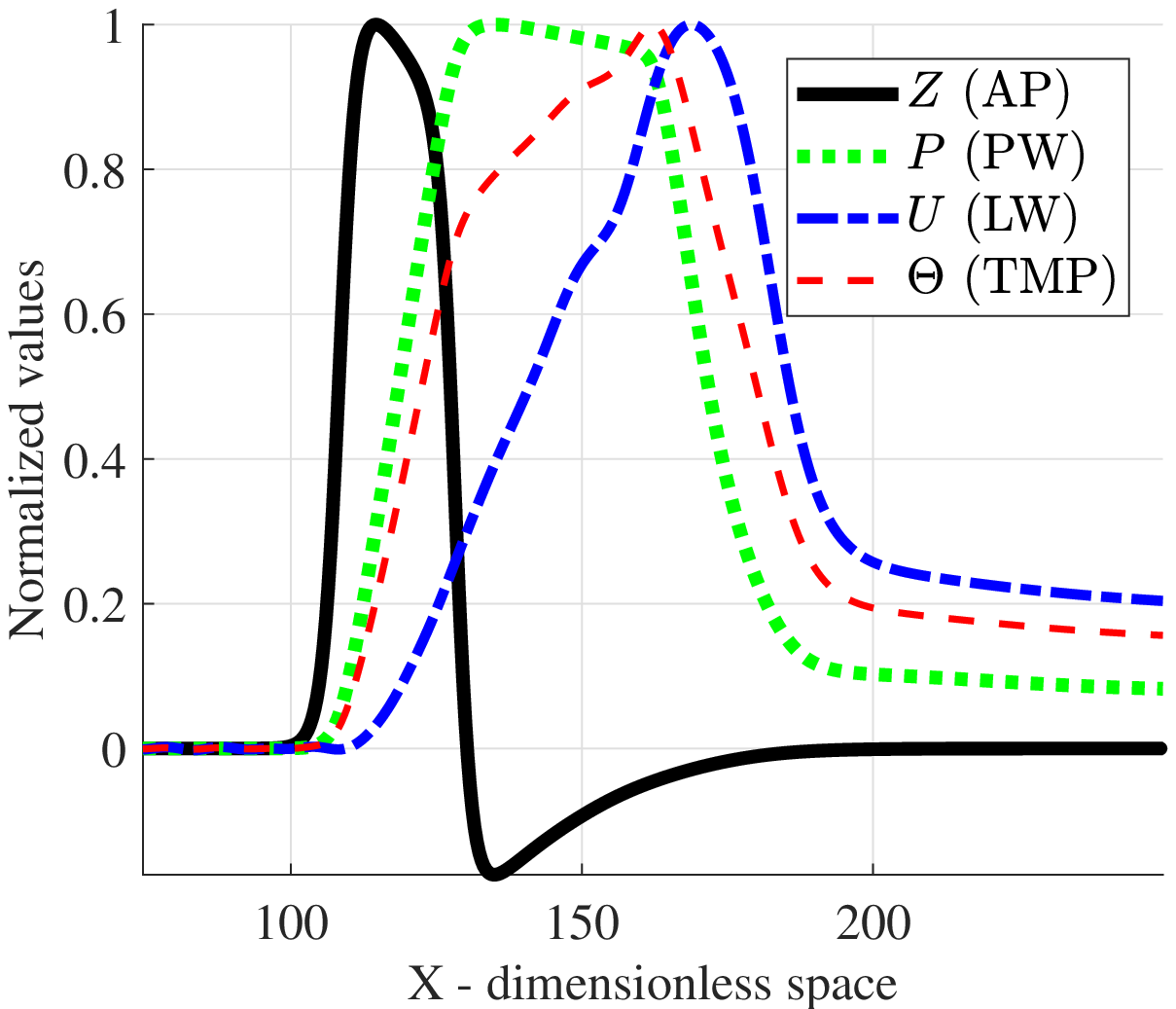}
\caption{Heat signals. Left panel (with small, but noticeable viscous dampening in $P$ and $U$, thermal sources from $\propto J^2$ and $\propto \varphi_2(P)$ and $\propto \varphi_3(U)$) -- $\tau_1=0.0167$, $\tau_2=0.0333$, $\tau_3=0.1667$, right panel (heat source only from $P_T$ and $U_T$ with negligible viscosity) -- $\tau_1=0$, $\tau_2=0.1$, $\tau_3=0.025$, $\mu_1=0.0001$, $\mu_2=0.0001$}\label{Fig3}
\end{figure}

Two distinct thermal response cases are shown in Fig.~\ref{Fig3}. It must be noted that the PW and LW profiles are also significantly different.  A case with full $F_3$ in the form of $\tau_1 J^{2} + \tau_2 \left( P_T + \varphi_2(P) \right) + \tau_3 \left( U_T + \varphi_3(U) \right) - \tau_4 K$ and the viscosity in $P$ \eqref{peq} and $U$ \eqref{ihjeq} is shown in left panel in Fig.~\ref{Fig3}. In the right panel in Fig.~\ref{Fig3} the irreversible thermal processes have been significantly suppressed by setting coefficient $\tau_1$ to zero (related to Joule heating) and reducing the viscosity by two orders of magnitude compared to the case in the left panel. This means that the thermal response in Fig.~\ref{Fig3} right panel is dominated by the reversible processes (temperature increase when density increases in axoplasm and membrane and temperature decrease if the density decreases with friction related terms suppressed). It can also be noted that viscosity or lack of it has a significant effect on the shape of the PW and LW profiles under the used parameters. The governing equations for the PW and LW are conservative before adding the contact forces $F_1$, $F_2$ and the dissipative terms. Without accounting for some kind of friction-like term in PW and LW the different sound velocities in the corresponding environments have an opportunity to spread the wave-profiles out as the driving force from the AP is taken with a different velocity than the sound speed in axoplasm and biomembrane. While not demonstrated in the figures, it should be noted the thermal signal drop off rate is dependent on the relaxation time $\tau_K$.
If relaxation is taken fast enough it is possible to model thermal response curves which are almost perfectly in-phase with the propagating wave ensemble.
Some endothermic chemical processes in the context of the nerve pulse propagation have been discussed by Abbott et al.~\citep{Abbott1958}.
Alternatively thermal response in-phase with the propagating wave ensemble can emerge if the reversible thermal changes are dominating (Fig.~\ref{Fig3} right panel).

\begin{figure}
\includegraphics[width=0.33\textwidth]{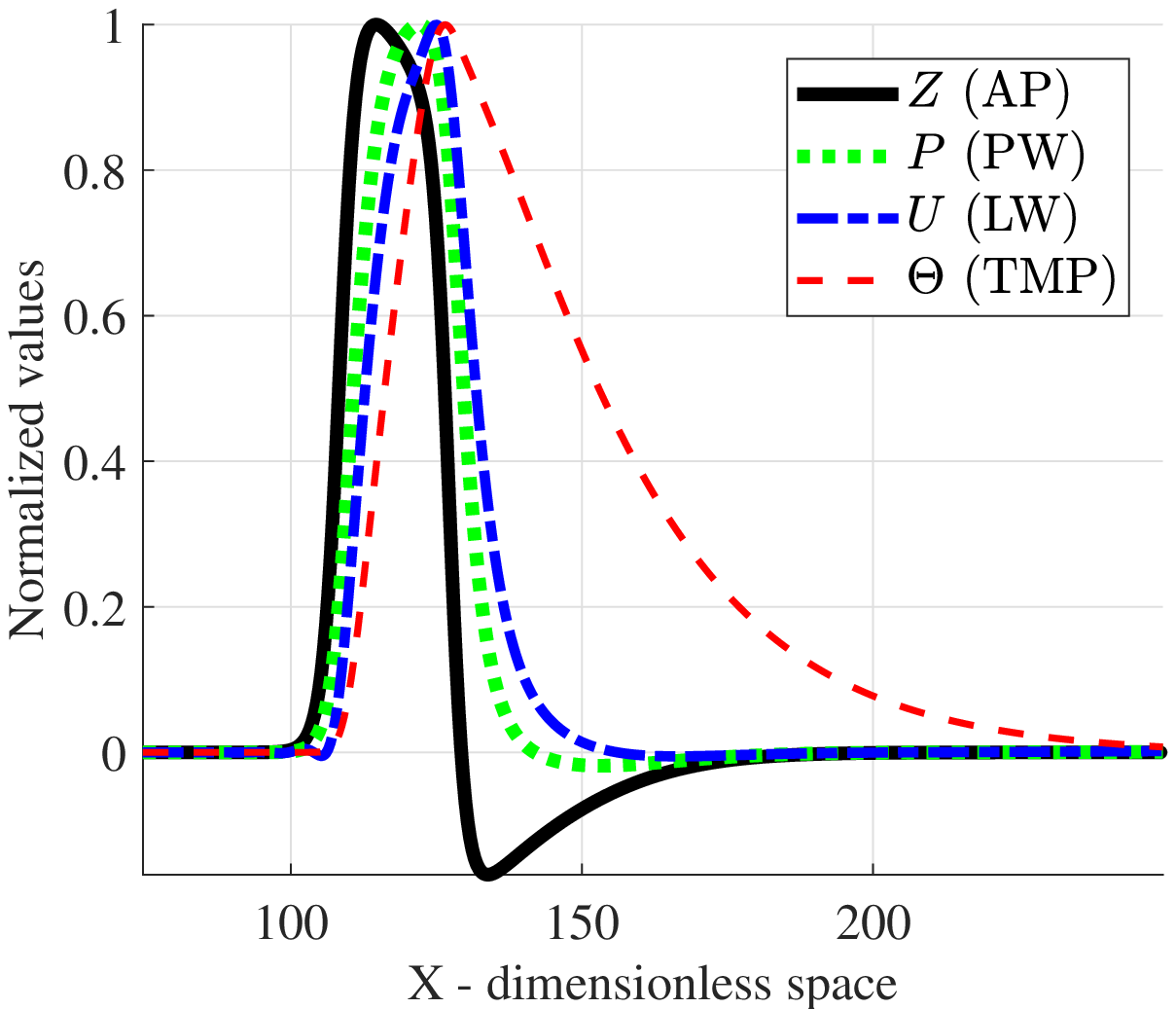}
\includegraphics[width=0.315\textwidth]{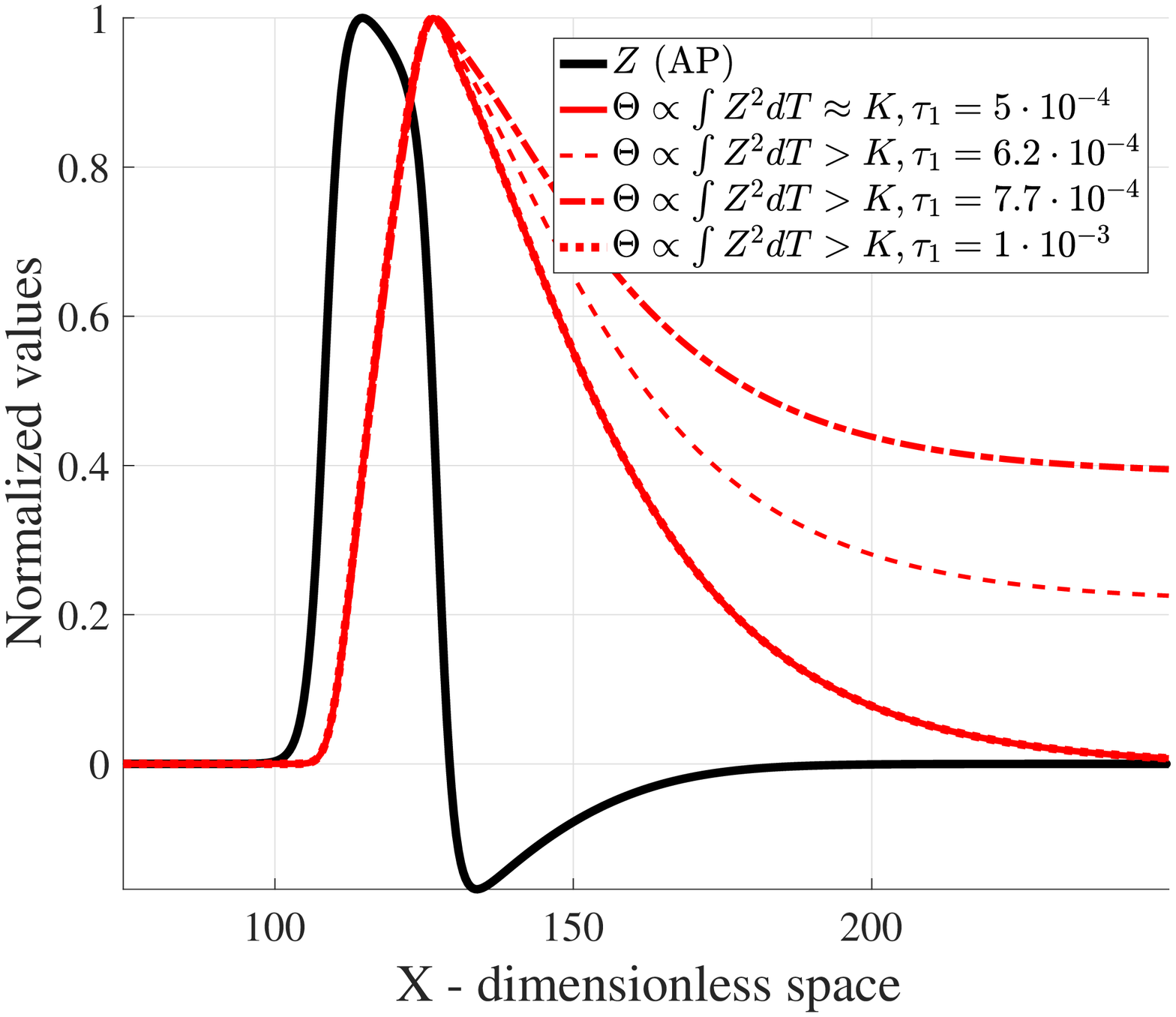}
\includegraphics[width=0.315\textwidth]{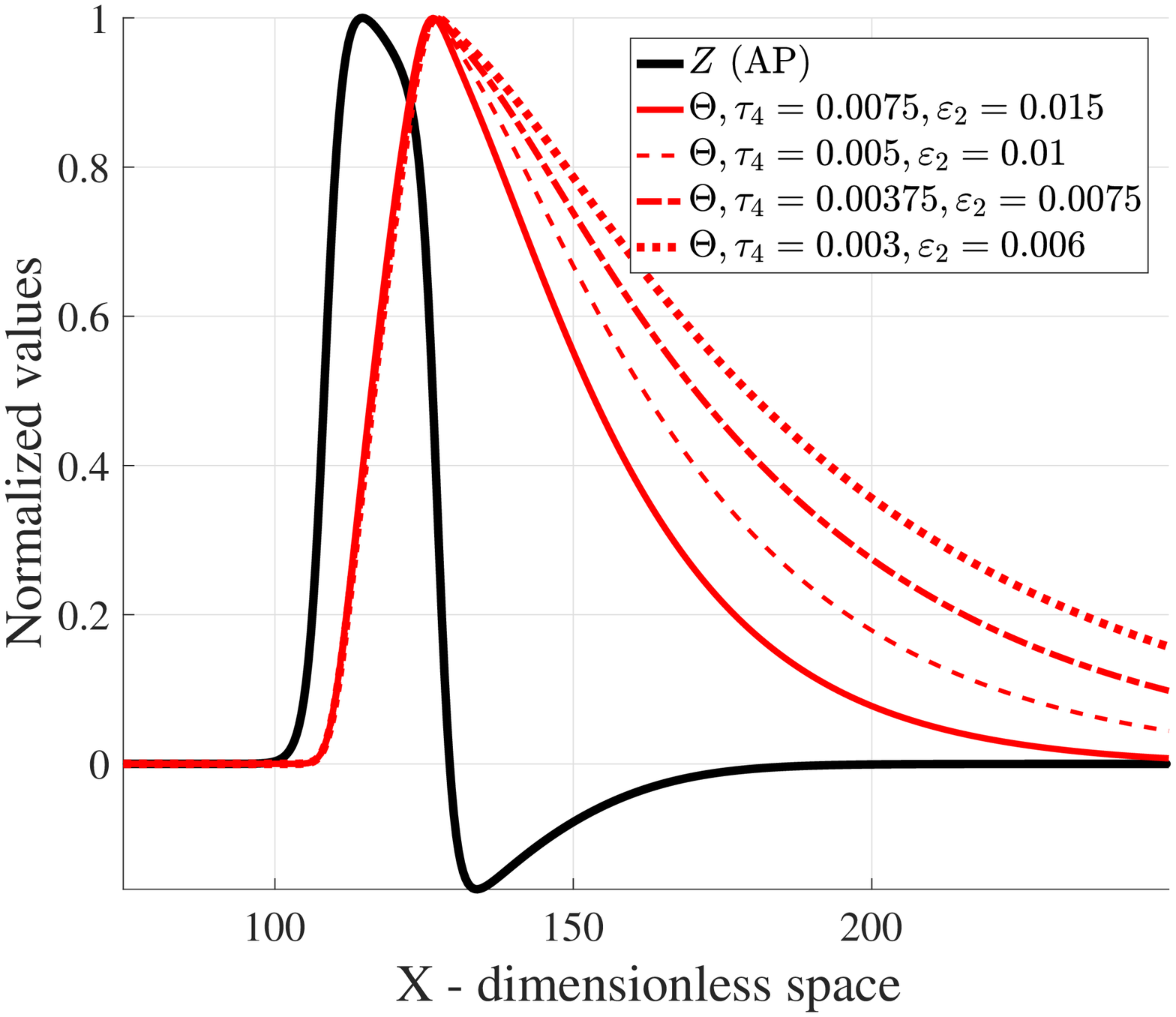}
\caption{Heat signals. Left panel (thermal source only $\propto Z^2$) balanced parameters so that $\tau_1 Z^2 \approx K$ the parameters are $\tau_1=0.0005$, $\tau_2=0$, $\tau_3=0$, middle panel (thermal source only $\propto Z^2$) unbalanced parameters so that $\tau_1 Z^2 > K$ the parameters are $\tau_1=0.001$, $\tau_2=0$, $\tau_3=0$, right panel (thermal source only $\propto Z^2$) the effect relaxation times on the heat signal. Balanced parameters are chosen so that $\tau_1 Z^2 \approx K$. }
	\label{Fig5}
\end{figure}

Figure~\ref{Fig5} demonstrates the effect of Joule heating as a function of $Z^2$. In the left panel the thermal source and thermal sink have been balanced $\tau_1 Z^2 \approx K$ so that the thermal response settles eventually back to its starting value. In the middle panel the  balance between thermal source and thermal sink terms is varied, demonstrating that depending on the balance between these terms there is a different residual level of heat increase after the nerve pulse wave ensemble has passed.
In the right panel the thermal source and thermal sink are balanced and the relaxation time is varied. Demonstrating that depending on the relaxation time related parameters, the equilibrium or residual level can be reached at different rates.
In terms of qualitative thermal response curve characteristics there is no practical difference between formulating the Joule heating in terms of $Z^2$ or $J^2$. 

The numerical scheme used (with minor changes) for the example solutions is described in \cite{Engelbrecht2018c,Tamm2019} and references therein. The parameters used in the present numerical examples are the following. The FHN equation \eqref{fhneq}: $D = 1$, $a_i = 0.2$, $\beta_i = 0.025$, $\epsilon_1=0.018$. 
The pressure equation \eqref{peq}: $c_{2}^{2}=0.09$, $\mu_2=0.05$ and coupling force $F_2$ \eqref{f1eq}: $\eta_1 =10^{-3}$, $\eta_2=10^{-2}$, $\eta_3=10^{-3}$. 
The iHJ equation \eqref{ihjeq}: $N=-0.05$, $M=0.02$, $H_1 = 0.2$, $H_2 = 0.99$, $c_{3}^{2} = 0.1$ and coupling force $F_3$ \eqref{f2eq}: $\gamma_1 =10^{-2}$, $\gamma_2=10^{-3}$, $\gamma_3=10^{-5}$. Parameter $\mu_3$ in \eqref{ihjeq} is taken as $\mu_3=0.05$ unless noted otherwise. 
The improved heat equation \eqref{tempeq2}: $\alpha = 0.05$ and thermal source/sink function $F_4$ \eqref{f3eq} (or $\eqref{f3eqz}$): $\tau_4 = \epsilon_4 / 2 = 0.005$ and $\tau_{1,2,3}$ is variable parameter. In addition, in \eqref{f3PUint} $\lambda_2 = \mu_2 = 0.05$ and $\lambda_3 = \mu_3 = 0.05$ unless noted otherwise. In \eqref{keq} $\zeta = 0.005$ and $\epsilon_4=0.01$ unless noted otherwise. 
The parameters for the numerical scheme and initial conditions: $n=2^{11}$ (number of spatial grid points), $L=128$ (the length of spatial period is $L\cdot 2\pi$), $\Delta T = 1$ and $T_f=800$ (integration time in dimensionless units). For the initial condition $A_z=1.2$ (amplitude of the initial excitation for $Z$) and $B_z=1$ (width of initial sech$^2$-type excitation for $Z$) while all other initial conditions are taken zero ($P,U,\Theta$ at rest).

\section{Discussion}

What has been described above, is the modelling of a complex process of propagation of signals in nerve fibres at the interface of physics, mathematics and physiology \citep{Engelbrecht2019}. The physical background of wave motion is taken as a basic phenomenon, then processes formulated in the mathematical language and finally based on physiological understandings on the single effects coupled into a whole. In some sense we have followed A. Einstein \citep{Einstein1934} who said: ``\ldots mathematical construction enables us to discover the concepts and laws connecting them which give us the key to the understanding of the phenomena of Nature''.

An overview on experimental physical effects permits to describe the possible mechanisms of interaction between electrical and mechanical signals accompanied by temperature changes. Based on our hypothesis on the role of changes in electrical signals as sources for accompanying effects \citep{EngelbrechtMEDHYP}, the full list of possible factors are described and the coupling forces formulated. Compared with our previous studies, the number of possible coupling effects is enlarged and a new internal variable is introduced in order to account for possible endothermic reactions which according to experiments \citep{Abbott1958} could influence the temperature changes. Like the phenomenological variables in the HH model, the change of this internal variable involves a certain equilibrium level and the corresponding relaxation time. In order to justify the energy transduction, the dissipative term is included into the governing equation of the LW. Although we have followed the HH paradigm by starting the process by an electrical signal, the possible feedback is taken into account by coupling forces.

So in the full coupled model the primary governing equations which are related to wave motion have a physical basis derived from basic principles of continuum theory. The non-wave-like phenomena are governed either by diffusion equation (temperature change) or derived from the primary variables (this concerns the TW). The numerical results described in Section 5, are quantitatively similar to experimental profiles of changes (see references in Section~5).

The presented model is certainly robust and could be improved or modified in many aspects. As explained in Section~5, the FHN model has been used just for obtaining a qualitatively correct shape of an AP with an overshoot and the corresponding ion current. If the HH model with many physical constants would be used then there would be a possibility to specify the role of various ion currents. This could modify the generation of accompanying effects and may explain the retardation of refraction and also the influence of anaesthetics. In the proposed model the mechanical activation of ion channels in the FHN model is described phenomenologically by means of various coefficients. Such a proposal might also need refinements by specifying activation forces from lipids (influence from the biomembrane) or filaments (influence from the axoplasm) \citep{Martinac2018}. The LW is governed by iHJ equation which is an excellent description of a mechanical wave in a biomembrane \citep{Heimburg2005}. However, one should note that although this equation has a soliton-type solution, its generation from an input needs time \citep{Engelbrecht2017} and the time-scale might not be large enough for a real soliton to be emerged during the AP propagation. In addition, the existence of ion channels in biomembranes is a clear sign for inhomogeneities in the lipid bi-layer which could also influence the mechanical wave. This is again a challenge for further improvement of the model. The molecular effects related to ion movement, the influence of membrane proteins and the existence of filaments in the axoplasm may have a considerable effect to coupling forces. At this stage, the possible role of chemical reactions accompanying the whole process is taken into account by an internal variable related to possible temperature changes. 

It is clear that every theoretical model should be validated by experiments. The mathematical modelling described above may be used to uncover new mechanisms that experimental science has not yet encountered \citep{Wooley2005}. Such an approach is justified from the viewpoint of physical considerations which must be correlated with physiological phenomena. The possible influence of various coupling forces depending on changes of variables demonstrated above (Section~5) indicates the possibilities for interpretation of experiments. 

The explanations on coupling forces given above actually represent a rather full theoretical framework of possible interactions. Although there is a doubt whether it is possible to build up ``general unifying models'' \citep{Holland2019}, one cannot overlook the mathematical description of the physical background because the signal propagation in nerves is a complex electro-mechano-thermo-physiological process spiced with molecular effects resulting in an ensemble of waves. Surely the different models used so far for the description of single effects may have different levels of idealization but this is not a fundamental obstacle. Every attempt to understand this complexity is welcome and needs a close collaboration of theoreticians and experimentalists. In other words, it is a challenge to put the pieces back to the whole. The proposed model is still far from being ideal but forms a backbone in modelling based on physical considerations of wave motion expressed in the mathematical language.

\section*{Acknowledgment} This research was supported by the European Union through the European Regional Development Fund (Estonian Programme TK 124) and by the Estonian Research Council (IUT 33-24).

\bibliographystyle{elsarticle-num}

\begin{thebibliography}{10}
\expandafter\ifx\csname url\endcsname\relax
  \def\url#1{\texttt{#1}}\fi
\expandafter\ifx\csname urlprefix\endcsname\relax\def\urlprefix{URL }\fi
\expandafter\ifx\csname href\endcsname\relax
  \def\href#1#2{#2} \def\path#1{#1}\fi

\bibitem{Bishop1956}
G.H.~Bishop, {Natural history of the nerve impulse}, Physiol.~Rev.~36~(3)
  (1956) 376--399.
\newblock \href {https://doi.org/10.1152/physrev.1956.36.3.376}
  {\path{doi:10.1152/physrev.1956.36.3.376}}.

\bibitem{Nelson2004}
P.C.~Nelson, M.~Radosavljevic, S.~Bromberg, {Biological physics: energy,
  information, life}, Physics Today 57~(11) (2004) 63--64.
\newblock \href {https://doi.org/10.1063/1.1839381}
  {\path{doi:10.1063/1.1839381}}.

\bibitem{Iwasa1980}
K.~Iwasa, I.~Tasaki, R.~Gibbons, {Swelling of nerve fibers associated with
  action potentials}, Science 210~(4467) (1980) 338--339.
\newblock \href {https://doi.org/10.1126/science.7423196}
  {\path{doi:10.1126/science.7423196}}.

\bibitem{Tasaki1988}
I.~Tasaki, {A macromolecular approach to excitation phenomena: mechanical and
  thermal changes in nerve during excitation}, Physiol.~Chem.~Phys.~Med.~NMR 20
  (1988) 251--268.

\bibitem{Tasaki1989}
I.~Tasaki, K.~Kusano, P.M.~Byrne, {Rapid mechanical and thermal changes in the
  garfish olfactory nerve associated with a propagated impulse}, Biophys.~J.~55~(6) (1989) 1033--1040.

\bibitem{Yang2018}
Y.~Yang, X.W.~Liu, H.~Wang, H.~Yu, Y.~Guan, S.~Wang, N.~Tao, {Imaging action
  potential in single mammalian neurons by tracking the accompanying
  sub-nanometer mechanical motion}, ACS Nano (2018)
  \href {https://doi.org/10.1021/acsnano.8b00867}
  {\path{doi:10.1021/acsnano.8b00867}}.

\bibitem{Terakawa1985}
S.~Terakawa, {Potential-dependent variations of the intracellular pressure in
  the intracellularly perfused squid giant axon.}, J.~Physiol.~369~(1) (1985)
  229--248.
\newblock \href {https://doi.org/10.1113/jphysiol.1985.sp015898}
  {\path{doi:10.1113/jphysiol.1985.sp015898}}.

\bibitem{Ritchie1985}
J.M.~Ritchie, R.D.~Keynes, {The production and absorption of heat associated
  with electrical activity in nerve and electric organ}, Q.~Rev.~Biophys.~18~(4) (1985) 451--476.
\newblock \href {https://doi.org/10.1017/S0033583500005382}
  {\path{doi:10.1017/S0033583500005382}}.

\bibitem{Howarth1968}
J.V.~Howarth, R.D.~Keynes, J.M.~Ritchie, {The origin of the initial heat
  associated with a single impulse in mammalian non-myelinated nerve fibres},
  J.~Physiol.~194~(3) (1968) 745--793.
\newblock \href {https://doi.org/10.1113/jphysiol.1968.sp008434}
  {\path{doi:10.1113/jphysiol.1968.sp008434}}.

\bibitem{Engelbrecht2018d}
J.~Engelbrecht, T.~Peets, K.~Tamm, {Soliton trains in dispersive media}, Low
  Temp.~Physics/Fizika Nizk.~Temp. 44~(7) (2018) 887--892.
\newblock \href {https://doi.org/10.1063/1.5041436}
  {\path{doi:10.1063/1.5041436}}.

\bibitem{Tamm2019}
K.~Tamm, J.~Engelbrecht, T.~Peets, {Temperature changes accompanying signal
  propagation in axons}, J.~Non-Equilib.~Thermodyn.~(2019) 
	\href {https://doi.org/10.1515/jnet-2019-0012}
  {\path{doi:10.1515/jnet-2019-0012}}.

\bibitem{Heimburg2005}
T.~Heimburg, A.D.~Jackson, {On soliton propagation in biomembranes and
  nerves.}, Proc.~Natl.~Acad.~Sci.~USA 102~(28) (2005) 9790--9795.
\newblock \href {https://doi.org/10.1073/pnas.0503823102}
  {\path{doi:10.1073/pnas.0503823102}}.

\bibitem{Chen2019}
H.~Chen, D.~Garcia-Gonzalez, A.~J{\'{e}}rusalem, {Computational model of the
  mechanoelectrophysiological coupling in axons with application to
  neuromodulation}, Phys.~Rev.~E 99~(3) (2019) 032406.
\newblock \href {https://doi.org/10.1103/PhysRevE.99.032406}
  {\path{doi:10.1103/PhysRevE.99.032406}}.

\bibitem{Hady2014}
A.~{El Hady}, B.B.~Machta, {Mechanical surface waves accompany action
  potential propagation}, Nat.~Commun.~6(2015) 6697.
\newblock \href {https://doi.org/10.1038/ncomms7697}
  {\path{doi:10.1038/ncomms7697}}.

\bibitem{Hodgkin1964a}
A.L.~Hodgkin, {The Conduction of the Nervous Impulse}, Liverpool University
  Press, 1964.

\bibitem{Andersen2009}
S.S.L.~Andersen, A.D.~Jackson, T.~Heimburg, {Towards a thermodynamic theory
  of nerve pulse propagation.}, Prog.~Neurobiol.~88~(2) (2009) 104--113.
\newblock \href {https://doi.org/10.1016/j.pneurobio.2009.03.002}
  {\path{doi:10.1016/j.pneurobio.2009.03.002}}.

\bibitem{Drukarch2018}
B.~Drukarch, H.A.~Holland, M.~Velichkov, J.J.~Geurts, P.~Voorn, G.~Glas,
  H.W.~de~Regt, {Thinking about the nerve impulse: A critical analysis of the
  electricity-centered conception of nerve excitability}, Prog.~Neurobiol.~169
  (2018) 172--185.
\newblock \href {https://doi.org/10.1016/j.pneurobio.2018.06.009}
  {\path{doi:10.1016/j.pneurobio.2018.06.009}}.

\bibitem{EngelbrechtMEDHYP}
J.~Engelbrecht, K.~Tamm, T.~Peets, {Modeling of complex signals in nerve
  fibers}, Med.~Hypotheses 120 (2018) 90--95.
\newblock \href {https://doi.org/10.1016/j.mehy.2018.08.021}
  {\path{doi:10.1016/j.mehy.2018.08.021}}.

\bibitem{Engelbrecht2018e}
J.~Engelbrecht, T.~Peets, K.~Tamm, {Electromechanical coupling of waves in
  nerve fibres}, Biomech.~Model.~Mechanobiol.~17~(6) (2018) 1771--1783.
\newblock \href {https://doi.org/10.1007/s10237-018-1055-2}
  {\path{doi:10.1007/s10237-018-1055-2}}.

\bibitem{Debanne2011}
D.~Debanne, E.~Campanac, A.~Bialowas, E.~Carlier, G.~Alcaraz, {Axon
  physiology}, Physiol.~Rev.~91~(2) (2011) 555--602.
\newblock \href {https://doi.org/10.1152/physrev.00048.2009.}
  {\path{doi:10.1152/physrev.00048.2009.}}

\bibitem{Mueller2014}
J.K.~Mueller, W.J.~Tyler, {A quantitative overview of biophysical forces
  impinging on neural function.}, Phys.~Biol.~11~(5) (2014) 051001.
\newblock \href {https://doi.org/10.1088/1478-3975/11/5/051001}
  {\path{doi:10.1088/1478-3975/11/5/051001}}.

\bibitem{Hodgkin1952}
A.L.~Hodgkin, A.F.~Huxley, {A quantitative description of membrane current
  and its application to conduction and excitation in nerve}, J.~Physiol.~117~(4) (1952) 500--544.
\newblock \href {https://doi.org/10.1113/jphysiol.1952.sp004764}
  {\path{doi:10.1113/jphysiol.1952.sp004764}}.

\bibitem{Tasaki1992}
I.~Tasaki, P.M.~Byrne, {Heat production associated with a propagated impulse
  in bullfrog myelinated nerve fibers}, Jpn.~J.~Physiol.~42~(5) (1992)
  805--813.
\newblock \href {https://doi.org/10.2170/jjphysiol.42.805}
  {\path{doi:10.2170/jjphysiol.42.805}}.

\bibitem{Watanabe1986}
A.~Watanabe, {Mechanical, thermal, and optical changes of the nerve membrane
  associated with excitation}, Jpn.~J.~Physiol.~36 (1986) 625--643.
\newblock \href {https://doi.org/10.2170/jjphysiol.36.625}
  {\path{doi:10.2170/jjphysiol.36.625}}.

\bibitem{Porubov2003}
A.V.~Porubov, {Amplification of Nonlinear Strain Waves in Solids}, World
  Scientific, Singapore, 2003.

\bibitem{Gonzalez-Perez2016}
A.~Gonzalez-Perez, L.~Mosgaard, R.~Budvytyte, E.~Villagran-Vargas, A.D.~Jackson,
  T.~Heimburg, {Solitary electromechanical pulses in lobster neurons}, Biophys.~Chem.~216 (2016) 51--59.
\newblock \href {https://doi.org/10.1016/j.bpc.2016.06.005}
  {\path{doi:10.1016/j.bpc.2016.06.005}}.

\bibitem{Perez-Camacho2017}
M.I.~Perez-Camacho, J.~Ruiz-Suarez, {Propagation of a thermo-mechanical
  perturbation on a lipid membrane}, Soft Matter 13~(37) (2017) 6555--6561.
\newblock \href {http://arxiv.org/abs/1705.05811} {\path{arXiv:1705.05811}},
  \href {https://doi.org/10.1039/C7SM00978J} {\path{doi:10.1039/C7SM00978J}}.

\bibitem{Fillafer2018}
C.~Fillafer, M.~Mussel, J.~Muchowski, M.F.~Schneider, {Cell surface
  deformation during an action potential}, Biophys.~J.~114~(2) (2018) 410--418.
\newblock \href {https://doi.org/10.1016/j.bpj.2017.11.3776}
  {\path{doi:10.1016/j.bpj.2017.11.3776}}.

\bibitem{Appali2010}
R.~Appali, S.~Petersen, U.~{Van Rienen}, {A comparision of Hodgkin-Huxley and
  soliton neural theories}, Adv.~Radio Sci.~8 (2010) 75--79.
\newblock \href {https://doi.org/10.5194/ars-8-75-2010}
  {\path{doi:10.5194/ars-8-75-2010}}.

\bibitem{Meissner2018a}
S.T.~Meissner, {Proposed tests of the soliton wave model of action potentials,
  and of inducible lipid pores, and how non-electrical phenomena might be
  consistent with the Hodgkin-Huxley model}, arXiv:1808.07193 [physics.bio-ph]
  (2018).

\bibitem{Clay2005}
J.R.~Clay, {Axonal excitability revisited}, Prog.~Biophys.~Mol.~Biol.~88~(1)
  (2005) 59--90.
\newblock \href {https://doi.org/10.1016/j.pbiomolbio.2003.12.004}
  {\path{doi:10.1016/j.pbiomolbio.2003.12.004}}.

\bibitem{Courtemanche1998}
M.~Courtemanche, R.J.~Ramirez, S.~Nattel, {Ionic mechanisms underlying human
  atrial action potential properties: insights from a mathematical model}, Am.~J.~Physiol.~275~(1) (1998) H301--H321.

\bibitem{Bean2007}
B.P.~Bean, {The action potential in mammalian central neurons}, Nat.~Rev.~Neurosci.~8~(6) (2007) 451--65.
\newblock \href {https://doi.org/10.1038/nrn2148} {\path{doi:10.1038/nrn2148}}.

\bibitem{Ranade2015}
S.S.~Ranade, R.~Syeda, A.~Patapoutian, {Mechanically activated ion channels},
  Neuron 87~(6) (2015) 1162--1179.
\newblock \href {https://doi.org/10.1016/j.neuron.2015.08.032}
  {\path{doi:10.1016/j.neuron.2015.08.032}}.

\bibitem{Gross1983}
D.~Gross, W.S.~Williams, J.A.~Connor, {Theory of electromechanical effects in
  nerve}, Cell.~Mol.~Neurobiol.~3~(2) (1983) 89--111.
\newblock \href {https://doi.org/10.1007/BF00735275}
  {\path{doi:10.1007/BF00735275}}.

\bibitem{Petrov2006}
A.G.~Petrov, {Electricity and mechanics of biomembrane systems:
  Flexoelectricity in living membranes}, Anal.~Chim.~Acta 568~(1-2) (2006)
  70--83.
\newblock \href {https://doi.org/10.1016/j.aca.2006.01.108}
  {\path{doi:10.1016/j.aca.2006.01.108}}.

\bibitem{Rvachev2010}
M.M.~Rvachev, {On axoplasmic pressure waves and their possible role in nerve
  impulse propagation}, Biophys.~Rev.~Lett.~5~(2) (2010) 73--88.
\newblock \href {https://doi.org/10.1142/S1793048010001147}
  {\path{doi:10.1142/S1793048010001147}}.

\bibitem{Barz2013}
H.~Barz, A.~Schreiber, U.~Barz, {Impulses and pressure waves cause excitement
  and conduction in the nervous system}, Med.~Hypotheses 81~(5) (2013)
  768--72.
\newblock \href {https://doi.org/10.1016/j.mehy.2013.07.049}
  {\path{doi:10.1016/j.mehy.2013.07.049}}.

\bibitem{Abbott1958}
B.C.~Abbott, A.V.~Hill, J.V.~Howarth, {The positive and negative heat
  production associated with a nerve impulse}, Proc.~R.~Soc.~B Biol.~Sci.~148~(931) (1958) 149--187.
\newblock \href {https://doi.org/10.1098/rspb.1958.0012}
  {\path{doi:10.1098/rspb.1958.0012}}.

\bibitem{Eringen1962}
A.C.~Eringen, {Nonlinear Theory of Continuous Media}, McGraw-Hill Book
  Company, New York, 1962.

\bibitem{Engelbrecht2015b}
J.~Engelbrecht, {Questions About Elastic Waves}, Springer International
  Publishing, Cham, 2015.
\newblock \href {https://doi.org/10.1007/978-3-319-14791-8}
  {\path{doi:10.1007/978-3-319-14791-8}}.

\bibitem{Hall1999}
C.W.~Hall, {Laws and Models: Science, Engineering, and Technology}, CRC Press,
  Boca Raton, 1999.

\bibitem{Engelbrecht2018arXiv}
J.~Engelbrecht, K.~Tamm, T.~Peets, {Primary and secondary components of nerve
  signals}, arXiv:1812.05335 [physics.bio-ph] (2018).

\bibitem{Engelbrecht2018c}
J.~Engelbrecht, T.~Peets, K.~Tamm, {Electromechanical coupling of waves in
  nerve fibres}, Biomech.~Model.~Mechanobiol.~17~(6) (2018) 1771--1783.
\newblock \href {http://arxiv.org/abs/1802.07014} {\path{arXiv:1802.07014}},
  \href {https://doi.org/10.1007/s10237-018-1055-2}
  {\path{doi:10.1007/s10237-018-1055-2}}.

\bibitem{Holland2019}
L.~Holland, H.W.~de~Regt, B.~Drukarch, {Thinking about the nerve impulse: the
  prospects for the development of a comprehensive account of nerve impulse
  propagation}, Front.~Cell.~Neurosci.~13 (2019).
\newblock \href {https://doi.org/10.3389/fncel.2019.00208}
  {\path{doi:10.3389/fncel.2019.00208}}.

\bibitem{Wooley2005}
{National Research Council}, {Catalyzing Inquiry at the Interface of Computing
  and Biology}, The National Academies Press, Washington, 2005.
\newblock \href {https://doi.org/10.17226/11480} {\path{doi:10.17226/11480}}.

\bibitem{Noble2010}
D.~Noble, {Biophysics and systems biology}, Philos.~Trans.~R.~Soc.~A Math.~Phys.~Eng.~Sci.~368~(1914) (2010) 1125--1139.
\newblock \href {https://doi.org/10.1098/rsta.2009.0245}
  {\path{doi:10.1098/rsta.2009.0245}}.
	
\bibitem{HJ2007}
T.~Heimburg, A.D.~Jackson, {On the action potential as a propagating density
  pulse and the role of anesthetics}, Biophysical Reviews and Letters 02~(01)
  (2007) 57--78.
\newblock \href {https://doi.org/10.1142/S179304800700043X}
  {\path{doi:10.1142/S179304800700043X}}.

\bibitem{Schneider2018}
M.~Mussel, M.F.~Schneider, {It sounds
  like an action potential: unification of electrical, chemical and mechanical
  aspects of acoustic pulses in lipids}, arXiv:1806.08551 [physics.bio-ph]
  (2018). 
\newblock \href {http://arxiv.org/abs/1806.08551}
	{\path{http://arxiv.org/abs/1806.08551}}.
	
\bibitem{Maugin1994a}
G.A.~Maugin, W.~Muschik, {Thermodynamics with internal variables part i.
  general concepts}, J.~Non-Equilib.~Thermodyn.~19~(3) (1994) 217--249.
\newblock \href {https://doi.org/10.1515/jnet.1994.19.3.217}
  {\path{doi:10.1515/jnet.1994.19.3.217}}.
	
\bibitem{Heimburg2008}
T.~Heimburg, A.D.~Jackson, {Thermodynamics of the nervous impulse}, in:
  N.~Kaushik (Ed.), Structure and Dynamics of Membranous Interfaces, John Wiley
  {\&} Sons, 2008, Ch.~12, 318--337.

\bibitem{Berezovski2017}
A.~Berezovski, P.~V{\'{a}}n, {Internal Variables in Thermoelasticity}, Springer, 2017.
\newblock \href {https://doi.org/10.1007/978-3-319-56934-5}
  {\path{doi:10.1007/978-3-319-56934-5}}.

\bibitem{Maugin1994}
G.A.~Maugin, J.~Engelbrecht, {A thermodynamical viewpoint on nerve pulse
  dynamics}, J.~Non-Equilib.~Thermodyn.~19~(1) (1994).
\newblock \href {https://doi.org/10.1515/jnet.1994.19.1.9}
  {\path{doi:10.1515/jnet.1994.19.1.9}}.

\bibitem{Van2008a}
P.~V{\'{a}}n, A.~Berezovski, J.~Engelbrecht, {Internal variables and dynamic
  degrees of freedom}, J.~Non-Equilib.~Thermodyn.~33~(3) (2008) 235--254.
\newblock \href {https://doi.org/10.1515/JNETDY.2008.010}
  {\path{doi:10.1515/JNETDY.2008.010}}.

\bibitem{Engelbrecht2019}
J.~Engelbrecht, K.~Tamm, T.~Peets, {Modelling of processes in nerve fibres at
  the interface of physiology and mathematics}, arXiv 1906.01261
  [physics.bio-ph] (2019).

\bibitem{Einstein1934}
A.~Einstein, {On the method of theoretical physics}, Philos.~Sci.~1~(2) (1934)
  163--169.

\bibitem{Martinac2018}
B.~Martinac, K.~Poole, {Mechanically activated ion channels}, Int.~J.~Biochem.~Cell Biol.~97 (2018) 104--107.
\newblock \href {https://doi.org/10.1016/j.biocel.2018.02.011}
  {\path{doi:10.1016/j.biocel.2018.02.011}}.

\bibitem{Engelbrecht2017}
J.~Engelbrecht, K.~Tamm, T.~Peets, {On solutions of a Boussinesq-type equation
  with displacement-dependent nonlinearities: the case of biomembranes},
  Philos.~Mag.~97~(12) (2017) 967--987.
\newblock \href {https://doi.org/10.1080/14786435.2017.1283070}
  {\path{doi:10.1080/14786435.2017.1283070}}.

\end{thebibliography}

\end{document}